\documentclass[aps,prc,twocolumn,showpacs,superscriptaddress,groupedaddress]{revtex4}

\usepackage{graphicx}
\usepackage{color}

\begin{document}

\preprint{1}

\title{Nonflow contribution to Dihadron Azimuthal Correlations in 200 GeV/c Au+Au Collisions}

\author{Yuhui Zhu}
\affiliation{Shanghai Institute of Applied Physics, Chinese Academy of Sciences, Shanghai 201800, China}
\affiliation{University of Chinese Academy of Sciences, Beijing 100049, China}

\author{Y. G. Ma\footnote{Author to whom all correspondence should be addressed:
ygma@sinap.ac.cn}}
\author{J.  H. Chen}
\author{G. L. Ma}
\author{S. Zhang}
\author{C. Zhong}

\affiliation{Shanghai Institute of Applied Physics, Chinese Academy of Sciences, Shanghai 201800, China}


\date{\today}

\begin{abstract}
Dihadron azimuthal correlations in 200 GeV/c Au+Au have been
simulated by a multi-phase transport (AMPT) model. Contribution
from jet-medium interaction to correlation function is obtained by
subtracting the combined harmonic flow background from the raw
dihadron correlation function. Signals in centralities of 0-10\%,
20-40\% and 50-80\% are compared in three associated transverse
momentum ($p_{T}^{assoc}$) bins: 0.2-0.8 GeV/c, 0.8-1.4 GeV/c and
1.4-2.0 GeV/c. An obvious medium modification impact can be seen
from the signal shape change and relative jet contribution in the
above events,  which shows different behaviors between central and
peripheral events, and among different $p_{T}^{assoc}$ ranges in
central events. More detailed $p_{T}^{assoc}$ dependence of the
derived nonflow contribution is studied in central 0-10\% events,
in which a strong $p_{T}^{assoc}$ dependence of RMS width is
observed. We also calculated that relative jet contributions in
peripheral and central collisions in the above mentioned cases.

 \end{abstract}
 \pacs{25.75.Gz, 12.38.Mh, 24.85.+p}

 \maketitle

 \section{Introduction}
 Lattice QCD calculations predicted a phase transition from a hadron
 gas to a deconfined matter in ultra-relativistic heavy ion
collisions~\cite{QCD-phasetran,White-papers}. A hot and
dense partonic matter formed in the process, called Quark-Gluon
Plasma (QGP),  is found to be strongly interacting experimentally.

Searching for the phase boundary and critical
point~\cite{QCD-phasetran,QCD,phase-th,Nu,PBM} has been always an important  topic in our physics world. For
this purpose, RHIC STAR Collaboration  completed a beam energy scan program
in 2010, which offers us a good way to vary the chemical potential
and temperature in the phase diagram.

For these years, an away-side double peak structure observed in RHIC experiments
has long been interpreted as the interaction between jets and medium,
therefore it is regarded as a signal of the QGP phase formation.
There have been many theoretical works on explaining the physical
mechanisms of this double peak structure, such as shock wave
model~\cite{Casalderrey},  gluon radiation model~\cite{Koch},
medium-induced gluon
bremsstrahlung~\cite{large-angle,opaque-media-radiation}, waking
the colored plasma and sonic Mach cones~\cite{Ruppert}, sonic
booms and diffusion wakes in thermal gauge-string
duality~\cite{sonic-booms}, jet deflection \cite{deflection} and
strong parton cascade mechanism
etc~\cite{ma-sqm,di-hadron,three-hadron,time-evolution,pt-dependence}.

However, constructing an ideal dihadron correlation background is
a complex task since it is contaminated by many uncertain sources.
For example, a higher harmonic flow background has been discussed
in several recent papers \cite{Ko,GL}. Their results show that
those odd orders of harmonic flows, such as triangular flow
($v_3$), which are induced by initial geometry fluctuations, can
significantly contribute to the away-side double peak structure.

In this paper, two methods for calculating background are employed and will be
discussed in detail. We concentrate on the study of transverse momentum ($p_{T}$) and centrality dependences of jet-medium contribution to the
dihadron azimuthal correlation functions. We investigate the two-particle away-side structure
in 200 GeV/c Au+Au collisions at different centralities of 0-10\%, 20-40\% and
50-80\%. We will focus on the central collisions where QGP is mostly
predicted to be produced, in which jet-induced signals in different associated $p_T$ ($p_{T}^{assoc}$) bins are shown.

The paper is organized as followed. Section II gives a
brief introduction on our simulation model. Section III describes our analysis method
for dihadron azimuthal correlations, especially for our background construction method
in detail. The
results and discussions about dihadron correlation are given in  Section  IV which is followed by 
a summary in  Section  V.

\section{Model Introduction}

In this paper, a hybrid model named as a multi-phase transport model
(AMPT)~\cite{AMPT}, is employed to study dihadron azimuthal
correlations. It includes four main components to describe the
physical processes in relativistic heavy-ion collisions: 1) the initial
conditions from HIJING model~\cite{HIJING}, 2) partonic
interactions modeled by Zhang's Parton Cascade model (ZPC)~\cite{ZPC},
3) hadronization, and 4) hadronic rescattering simulated by A
Relativistic Transport (ART) model~\cite{ART}.
The basic flow of simulation is in the following. First, many excited strings initiated from HIJING are melted into
partons in the AMPT version with string melting
mechanism~\cite{SAMPT}(abbr. ``the Melting AMPT version")
and a simple quark coalescence model is used to combine the
partons into hadrons. On the other hand, in the default version of AMPT
model~\cite{DAMPT}(abbr. ``the Default AMPT version"),
minijet partons are recombined with their parent strings when they
stop interactions and the resulting strings are converted to
hadrons via the Lund string fragmentation model~\cite{Lund}. Therefore, the
Melting AMPT version undergoes a partonic phase much more than
 the Default AMPT version. Details of the AMPT
model can be found in a review paper~\cite{AMPT} and previous
works~\cite{AMPT,SAMPT,Jinhui}.

We use the Melting AMPT version to do the
simulation for 200 GeV/c Au+Au collisions. In order to concentrate on partonic stage interactions, final
hadronic rescattering process is turned off in our simulation as well.

\section{Analysis Method}

The analysis method for raw dihadron azimuthal correlations is similar
to that used in previous
experiments~\cite{soft-soft-ex,sideward-peak2} which derives the
azimuthal correlation between a high $p_{T}$ particle (trigger
particle) and low $p_{T}$ particles (associated particles). In our
work, we give a $p_T >$ 2.5 GeV/c  cut on trigger particles and
select the associated particles whose $p_T$ is smaller than 2.5 GeV/c.
Both the trigger and
associated particles are required to be within a pseudo-rapidity
window of $|\eta|<1.0$. The raw signal is obtained by accumulating
pairs of trigger and associated particles into $\Delta\phi =
\phi_{assoc} - \phi_{trig}$ distributions in the same event.

\begin{figure*}[hbtp]
\includegraphics[width=5.2cm]{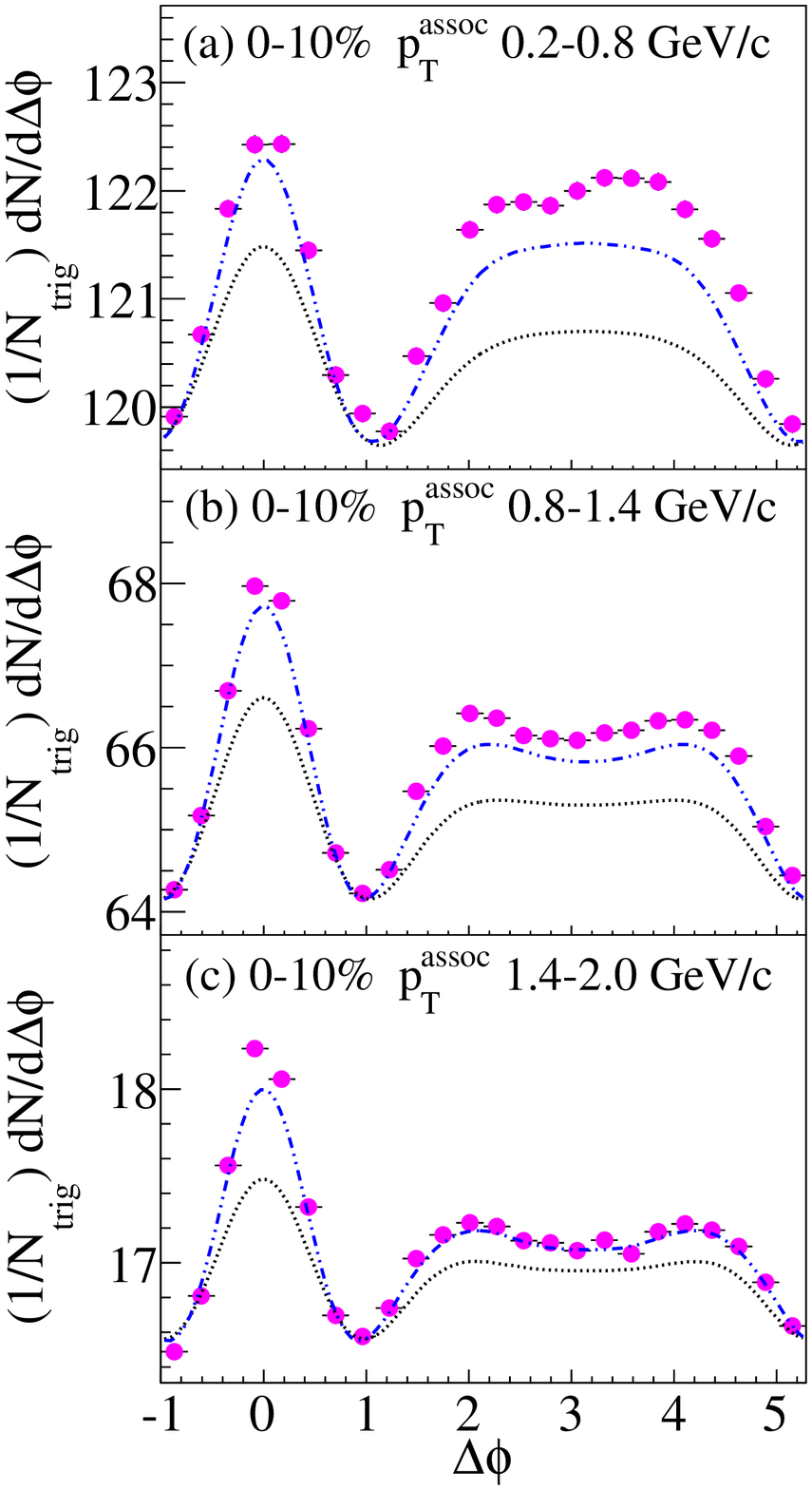}
\includegraphics[width=5.2cm]{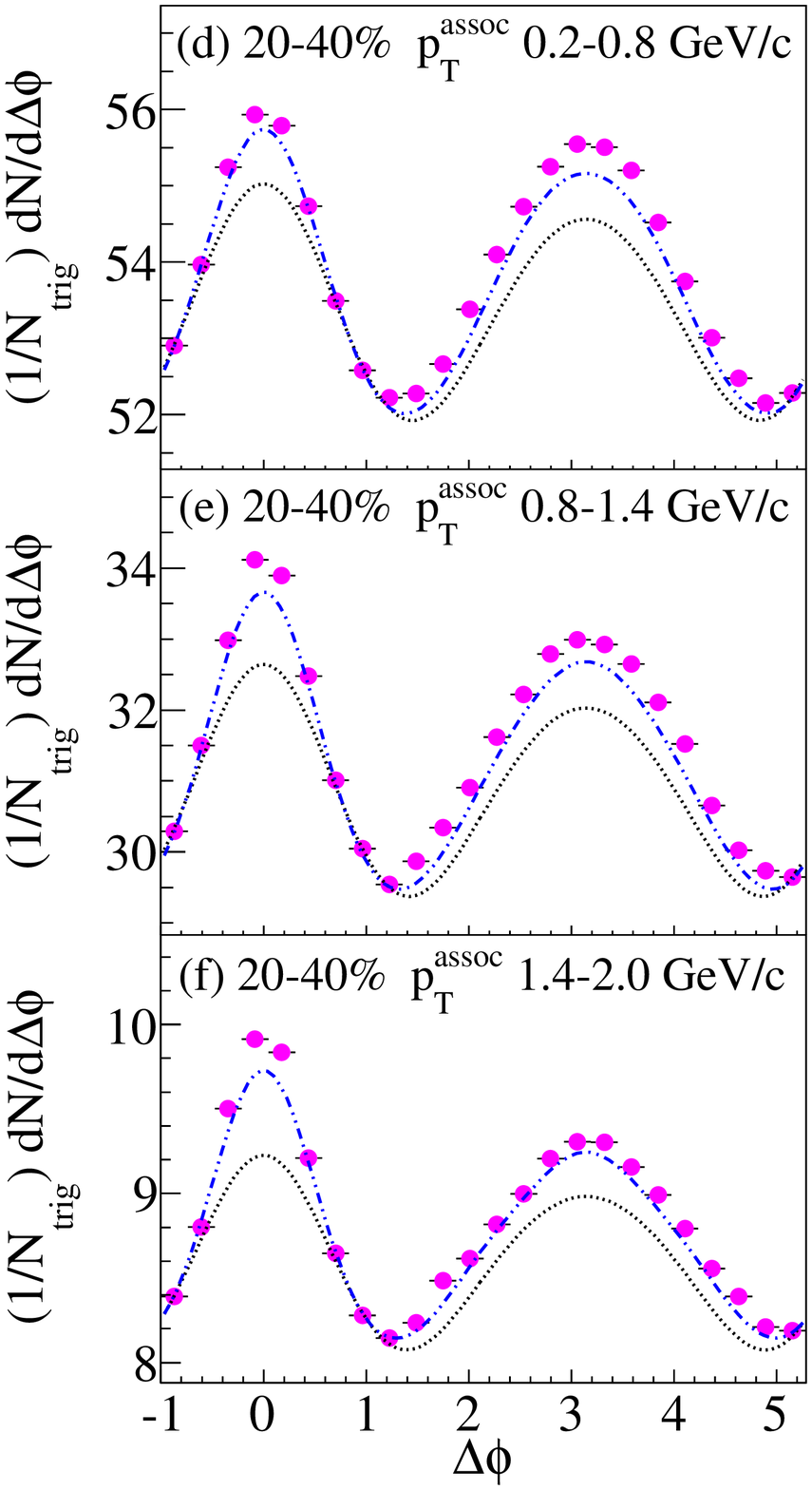}
\includegraphics[width=5.2cm]{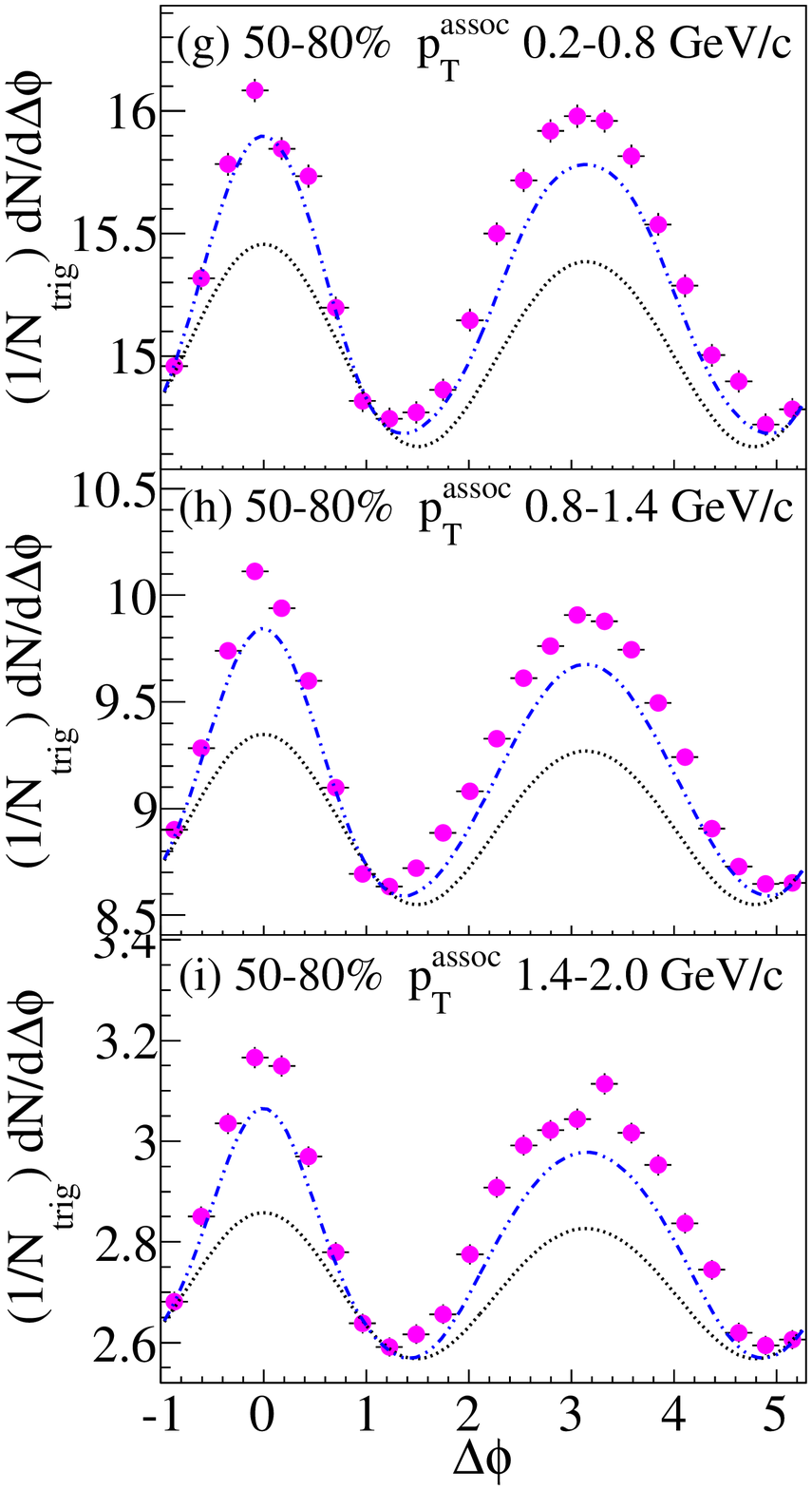}
\caption{(Color online) Dihadron correlation functions in 200 GeV/c
Au+Au collisions derived in three $p_{T}^{assoc}$ bins as illustrated in the
    plots.
 (Left)  0-10\% centrality,
 (Middle)   20-40\% centrality,
 (Right)  50-80\% centrality .
Solid circles are raw signals. Dash dotted blue lines are harmonic
combinational backgrounds constructed using formula (1) and dash black lines are
backgrounds constructed using formula (4).}
\label{200GeV_dihadron}
\end{figure*}

Normally the dihadron combinational background can be described by the formula of

\begin{widetext}
\begin{equation}
\left \langle f(\Delta \phi ) \right \rangle_{e}=\left \langle
\frac{N^{trig}\cdot N^{assoc}}{2\pi } \right \rangle_{e}+\left \langle
\frac{N^{trig}\cdot N^{assoc}}{2\pi }\cdot 2\sum_{n=1}^{+\infty}\upsilon
_{n}^{trig}\cdot \upsilon _{n}^{assoc} \right \rangle_{e}cosn\Delta \phi
\end{equation}
\end{widetext}
as used in Ref.~\cite{Ko,GL}, where suffix $e$ in $\left \langle ...\right \rangle_{e}$stands for
event-averaged quantity, quantities without suffixes are ones for each event.

We adopt the same analysis method as in Ref.~\cite{GL} and obtain the initial geometry
anisotropy $\upsilon_{n}^{r}$ with event plane angle from the
stage before the parton cascade, in order to exclude as much nonflow contribution as
possible. We first calculate the event plane angle
of each harmonic order by using the following formula:
\begin{equation}
\Psi_{n}^{r}=\frac{1}{n}\left [ arctan\frac{\left \langle r^{n}sin(n\phi)
    \right \rangle}{\left \langle r^{n}cos(n\phi ) \right \rangle} +\pi \right
    ],
    \end{equation}
where $r$ and $\phi$ are the polar coordinates of partons in the initial coordinate space.
Then we can obtain $\upsilon_{n}^{assoc}$ and $\upsilon_{n}^{trig}$ using the
formula:
\begin{equation}
\upsilon _{n}^{r}=\left \langle cos\left [ n(\phi -\Psi _{n}^{r}) \right ]
\right \rangle,
\end{equation}
where $\phi$ is the azimuthal angle in the final state momentum
space.  It effectively eliminates much nonflow contributions to
harmonic flows by this way \cite{Han}. From the $\upsilon_{n}$
versus $p_{T}$ plots as in Ref.~\cite{GL}, we can see that even in
the most central collision (b = 0 fm), it is better to include
higher order flow (up to fifth order) in background construction.
Therefore, we include up to fifth order flow contribution in
background reconstruction.

However, it was worried that if one multiplies
$\upsilon_{n}^{assoc}$ and $\upsilon_{n}^{trig}$ before
being event averaged, two-particle $\upsilon_{n}$ correlation could be a large contribution
to the background, i.e. whether the factorization ($\left \langle \upsilon_{n}^{assoc}  \upsilon_{n}^{trig} \right \rangle_{e}$= $\left \langle \upsilon_{n}^{assoc} \right \rangle_{e} \left \langle\upsilon_{n}^{trig} \right \rangle_{e}$) is held or not.  It has been noticed that the other modified form of the formula is also used in
some works \cite{STAR-dihadron-1,STAR-dihadron-2}(It should be noted
		here that they apply systematic errors on $\upsilon_{n}$, using background
		(1) as the upper bound):

\begin{widetext}
\begin{equation}
\left \langle f(\Delta \phi ) \right \rangle_{e}=\left \langle
\frac{N^{trig}\cdot N^{assoc}}{2\pi } \right \rangle_{e}+\left \langle
\frac{N^{trig}\cdot N^{assoc}}{2\pi } \right
\rangle_{e}2\sum_{n=1}^{+\infty}\left \langle \upsilon _{n}^{trig}\right
\rangle_{e}\left \langle \upsilon _{n}^{assoc} \right \rangle_{e}cosn\Delta\phi .
\end{equation}
\end{widetext}

The difference between formula (1) and (4) contains two kinds of contributions: 1. Flow
fluctuation and its correlation from initial geometry asymmetry;
2. nonflow and its correlation. The first kind of contribution
should be included in background while the second kind should be
excluded. However, neither background (1) and background (4) can
meet the demand. Background (1) overestimates background by
including nonflow contribution while background (4) underestimates
the background by throwing away the background from  flow
fluctuation and its correlation.  It is worth mentioning that
there have been some efforts to solve the crucial problem on
decomposing flow,
 flow fluctuation, and nonflow~\cite{purduenonflow}.

Although an ideal background is hard to be obtained, we can use the two
formulas (1) and (4)  as the upper and lower limits of the background to
get a reasonable range of jet-medium contribution. Next, background is reconstructed using formula (1) and (4)
as the upper and lower limits for dihadron background.
Correspondingly they are marked as ``background (1)" and ``background (4)" in
the figures. The detailed values of parameters $\left
\langle\upsilon_{1}^{trig}\right \rangle$, $\left
\langle\upsilon_{1}^{assoc}\right \rangle$ and $\left \langle\upsilon_{1}^{trig} \upsilon_{1}^{assoc} \right \rangle$ used in these two background
extractions are listed in appendix A.

From the formulas (1) and (4), we can see that the normalization
factor of the background should be $\left \langle
\frac{N^{trig}\cdot N^{assoc}}{2\pi } \right \rangle_{e}$
theoretically. However, only ZYAM scheme (A Zero Yield At Minimum)
can be used to adjust this factor to match the signal best
experimentally. However, we checked that the difference between
the theoretical normalization factor and  the ZYAM adjustment is
less than 3\%.

\begin{figure*}[hbtp]
\includegraphics[width=6.8cm]{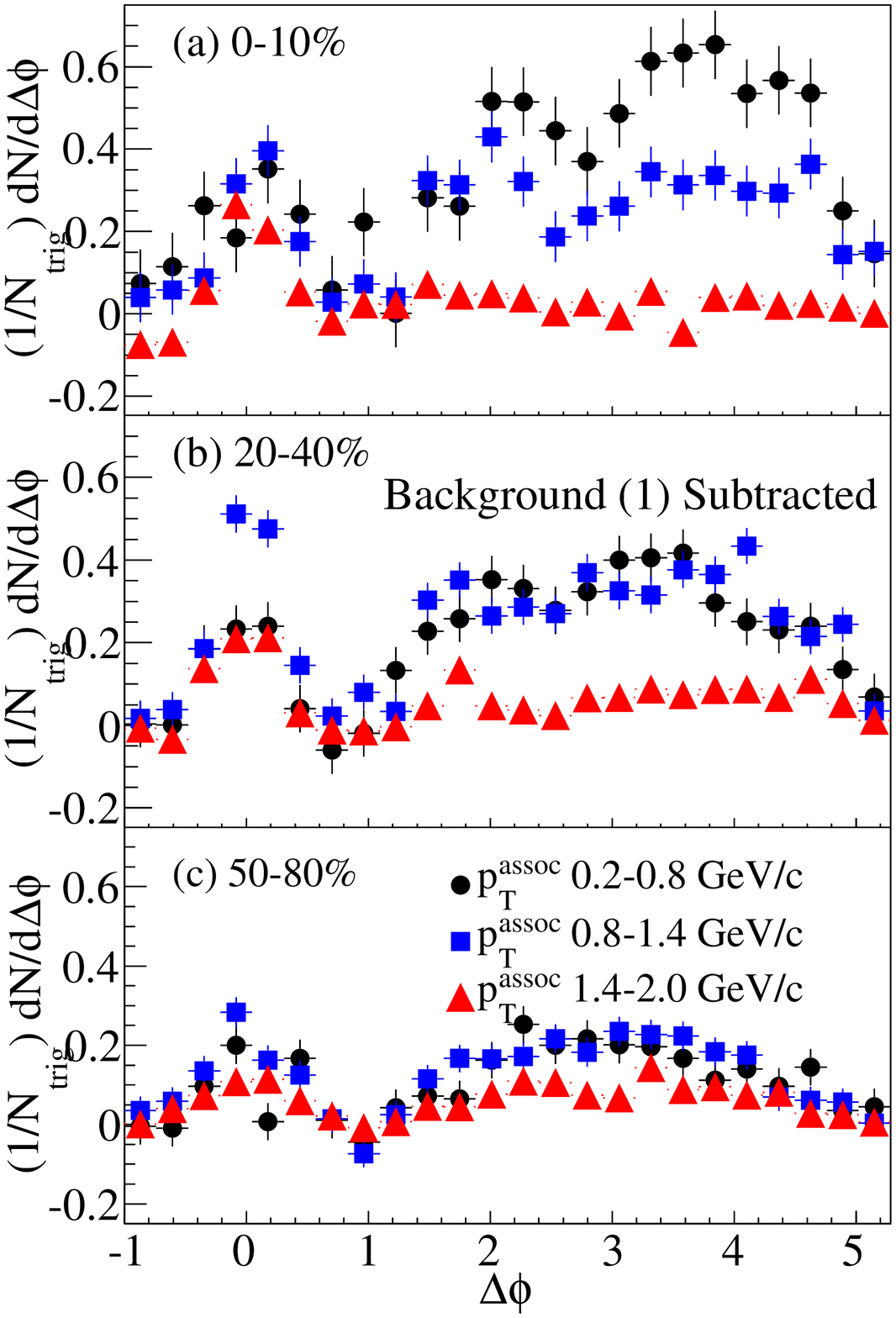}
\includegraphics[width=6.8cm]{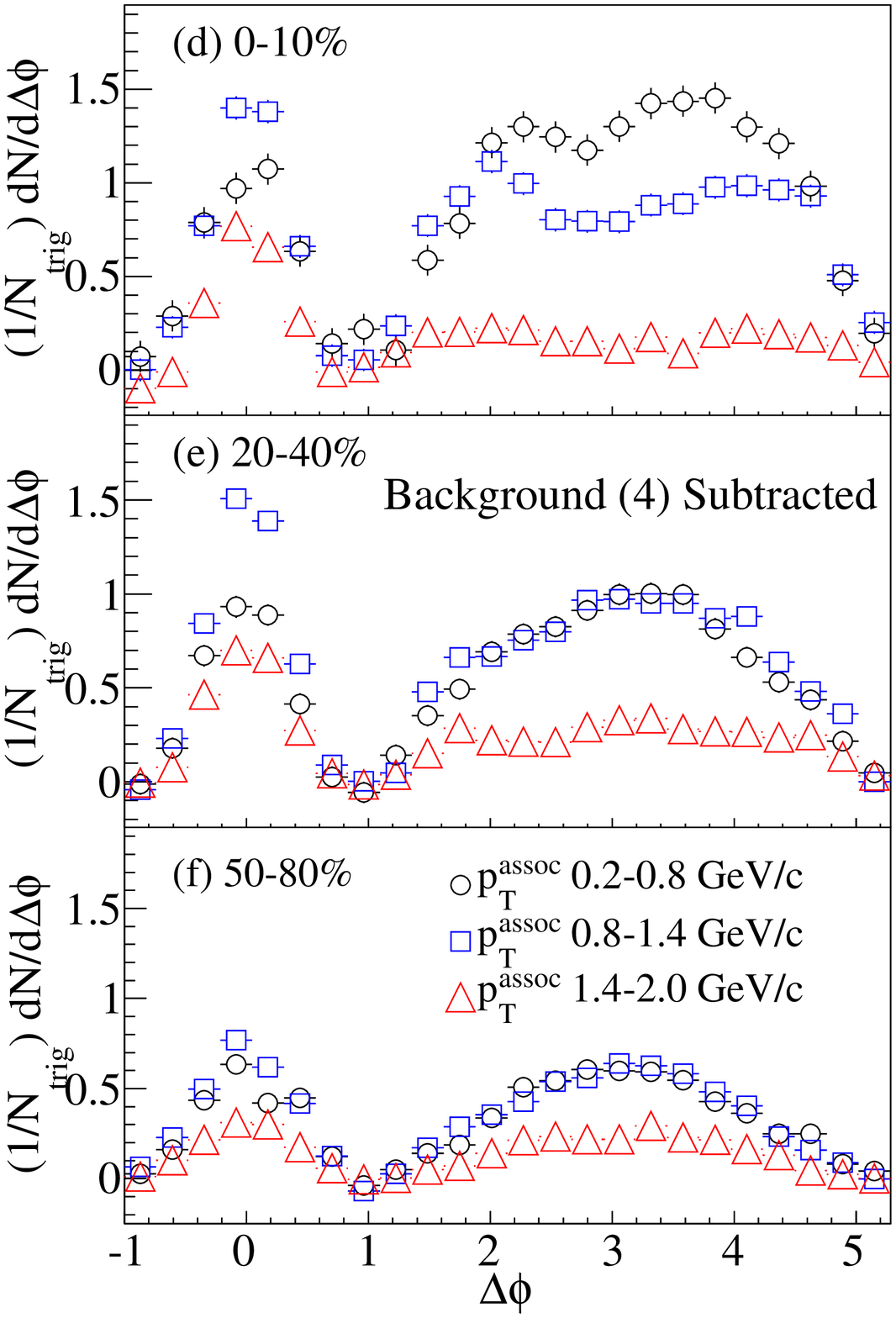}
\caption{(Color online) Background subtracted signals in three $p_{T}^{assoc}$ bins for three centralities.
Panels (a), (b), (c) stand for the signals with background (1) subtracted;
panels (d), (e), (f) stands for the signals after background (4) substracted. See texts for details.
}
\label{200GeV_dihadron_real_bg1}
\end{figure*}

\begin{figure*}[hbtp]
\includegraphics[width=6.8cm]{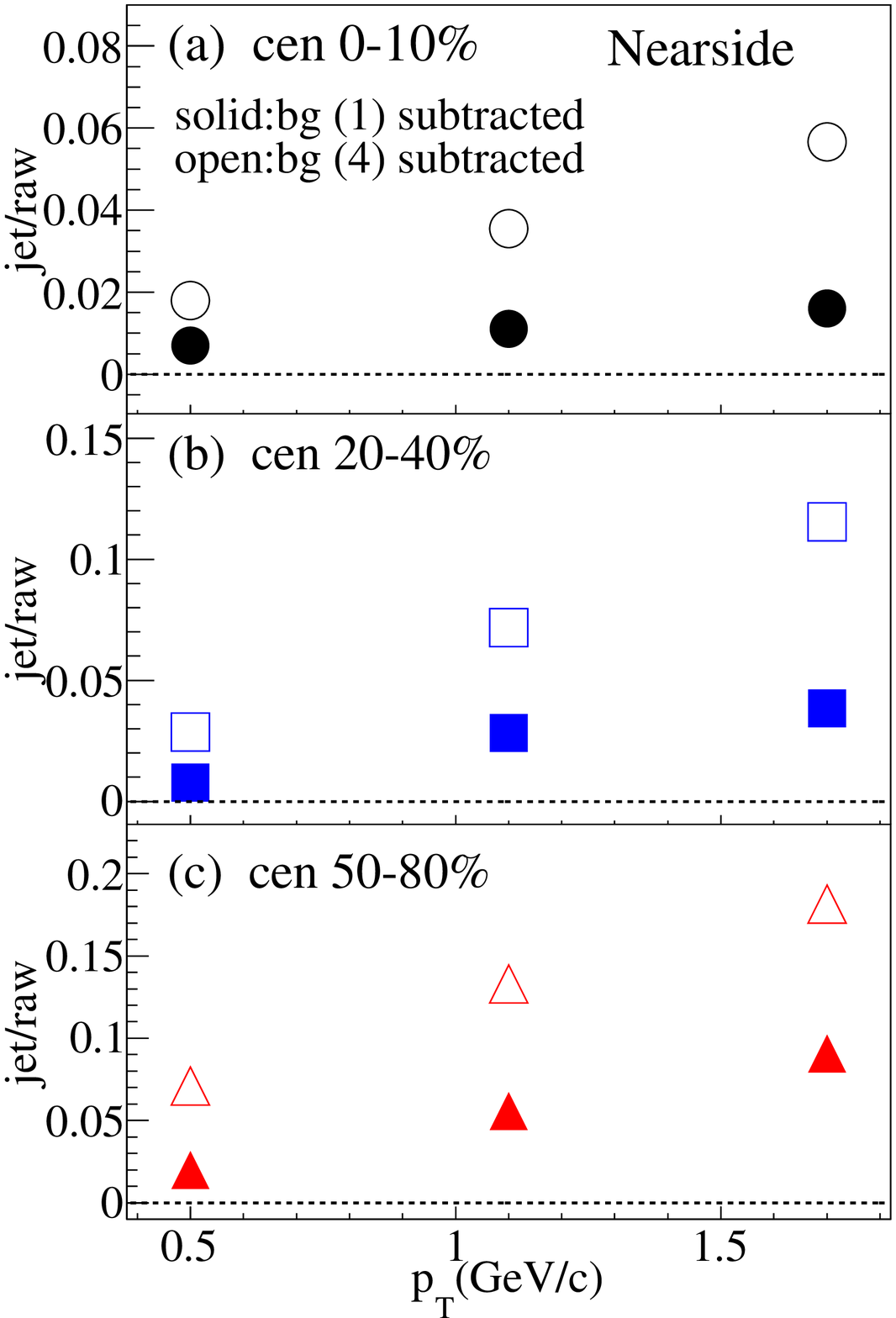}
\includegraphics[width=6.8cm]{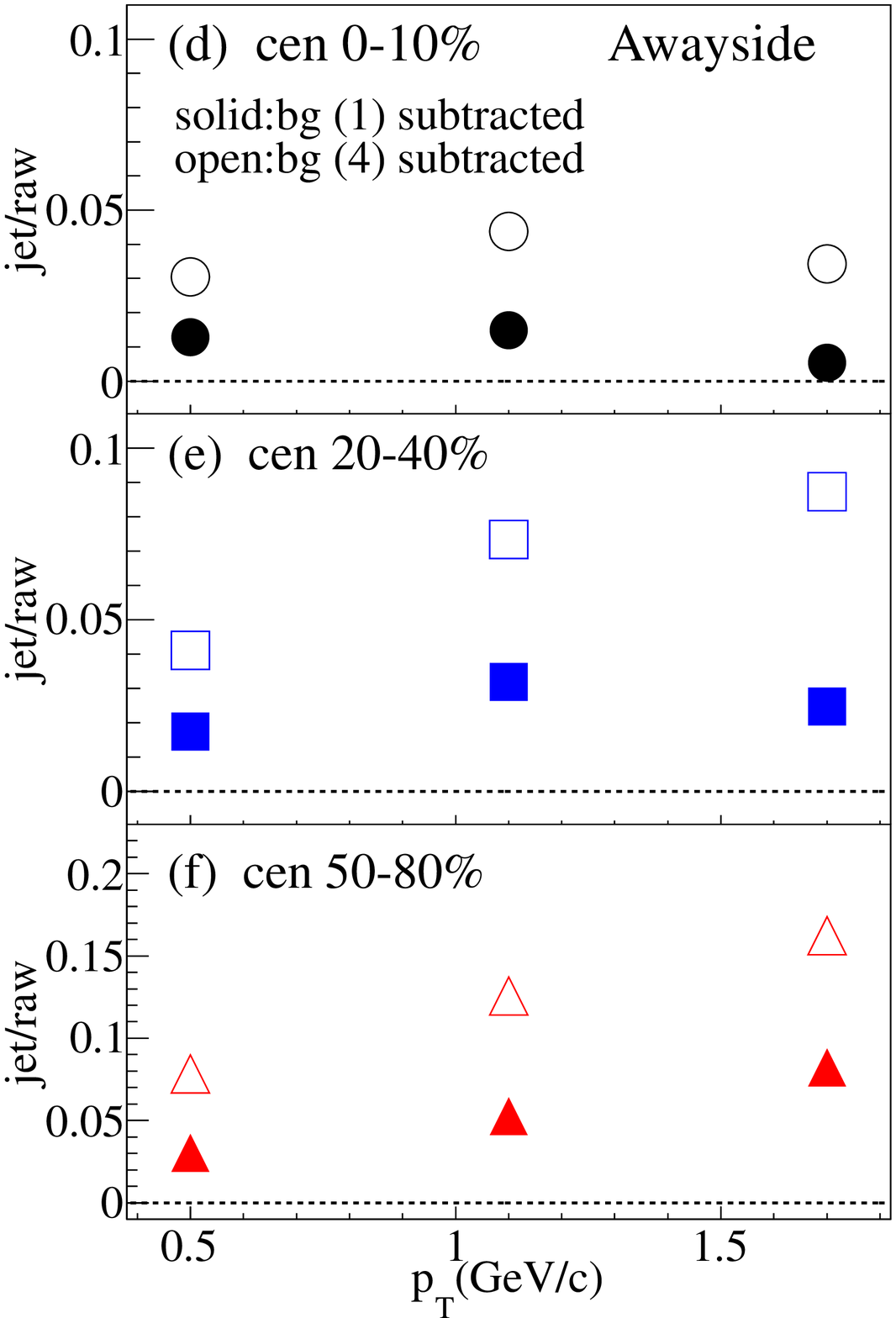}
\caption{(Color online) Jet relative contributions as a function of
	$p_{T}^{assoc}$ in three centrality bins for near-side (panels:(a),(b),(c)) and away-side (panels:(d),(e),(f)).
}
\label{200GeV_dihadron_realratio_near}
\end{figure*}

\section{Results and Discussions}

\subsection{Two Particle Azimuthal Correlation in Three Centrality Bins}

Fig. 1 shows the dihadron azimuthal correlations (both raw and combined
harmonic background) which are divided into three $p_{T}^{assoc}$ bins and
three centrality bins. Two backgrounds are drawn simultaneously as a
comparison.

First, we can observe an obvious difference in background shape in different
centralities. The central 0-10\% events have flatter or even double-peak shape
background, while 20-40\% and 50-80\% have a single-gaussian shape
background. That is due to the different propagation properties of harmonic flows for different centralities.
The existence of the hot dense matter (QGP) propagates the initial geometry
irregularities to a larger extent. In comparison to the tendency of $p_T$ dependence of $v_2$,
$v_3$ shows a stronger dependence especially in large $p_T$ range \cite{Han},
which makes $\upsilon_{3}$ increases more rapidly and further results in the
change of combinational background shape as one can
see the background shape becomes flatter and becomes a double peak
in the $p_{T}$ range from 1.4 to 2.0 GeV/c.  For the two backgrounds, they tend to have similar shapes and just differs a little bit in magnitudes.

Second, we can see a seemingly changing trend of the signal using either
background. Therefore, we need to subtract the background and further study the jet-medium contribution. The
corresponding plots are shown in Fig. 2. In Fig. 2, first from the global comparisons between different pads, we can observe an obvious change in signal
shape. Signals in more central events (0-10\%) or higher $p_{T}^{assoc}$
range (eg. 1.4-2.0 GeV/c) tend to have flatter signals while peripheral
50-80\% events almost show a single-peak shape signal. This difference is consistent with the medium modification picture. The
existence of QGP strongly modifies jets which makes the correlation shape
flatter (or more broadened) in more central collisions. In addition, it also suppresses higher
$p_{T}$ particles. On the contrary, if jets just interact less with surrounding medium particles, the correlation shape will tend to be single gaussian.
In addition, the previous results indicate that there should be some parts of contribution from hot spots by switching off jet production~\cite{GL}.

In this analysis, a quantity named ``jet relative contribution" is used to
represent the contribution of jet-medium correlation in total dihadron
correlation function. It is defined as the jet-medium correlation function yield divided by the raw
dihadron correlation yield (including flow background). These jet relative contributions
in different centrality bins and $p_{T}^{assoc}$ bins are obtained and shown
in Fig. 3.

Panels (a), (b), (c) in Fig. 3 are for near-side and Panels (d), (e), (f) are for
away-side. The jet relative contribution using two different
backgrounds are drawn together to provide an upper and lower
limit. From panel (d), one can see that the away-side jet relative
contribution in central 0-10\% events drops in $p_{T}^{assoc}$
range from 1.4 to 2.0 GeV/c, different from the case on near-side
(panel (a)). This is consistent with the high $p_{T}$ suppression
picture in QGP. The quantitative values are provided in Table I
and II. In general, the away-side jet relative contribution is
less than 5\% in central 0-10\% events.

\begin{table}
\caption{\label{tab:example}Near-side Jet-medium Contribution}
\begin{ruledtabular}
\begin{tabular}{llll}
   & 0.2-0.8 GeV/c & 0.8-1.4 GeV/c & 1.4-2.0 GeV/c\\
    0-10\% & 0.7\%-1.8\% & 1.1\%-3.5\% & 1.6\%-5.7\%\\
    20-40\% & 0.8\%-2.9\% & 2.8\%-7.2\% & 3.8\%-11.5\%\\
    50-80\% & 2.0\%-7.1\% & 5.2\%-13.3\% & 9.1\%-18.1\%\\
\end{tabular}
\end{ruledtabular}
\end{table}

\begin{table}
\caption{\label{tab:example2}Away-side Jet-medium Contribution}
\begin{ruledtabular}
\begin{tabular}{llll}
   & 0.2-0.8 GeV/c & 0.8-1.4 GeV/c & 1.4-2.0 GeV/c\\
    0-10\% & 1.3\%-3.1\% & 1.5\%-4.4\% & 0.5\%-3.4\%\\
    20-40\% & 1.7\%-4.1\% & 3.2\%-7.3\% & 2.5\%-8.7\%\\
    50-80\% & 3.0\%-7.8\% & 5.2\%-12.6\% & 8.2\%-16.2\%\\
\end{tabular}
\end{ruledtabular}
\end{table}

\subsection{$p_{T}^{assoc}$ dependence of jet-medium contribution in central 0-10\% collisions}

\begin{figure*}[hbtp]
\includegraphics[width=5.8cm]{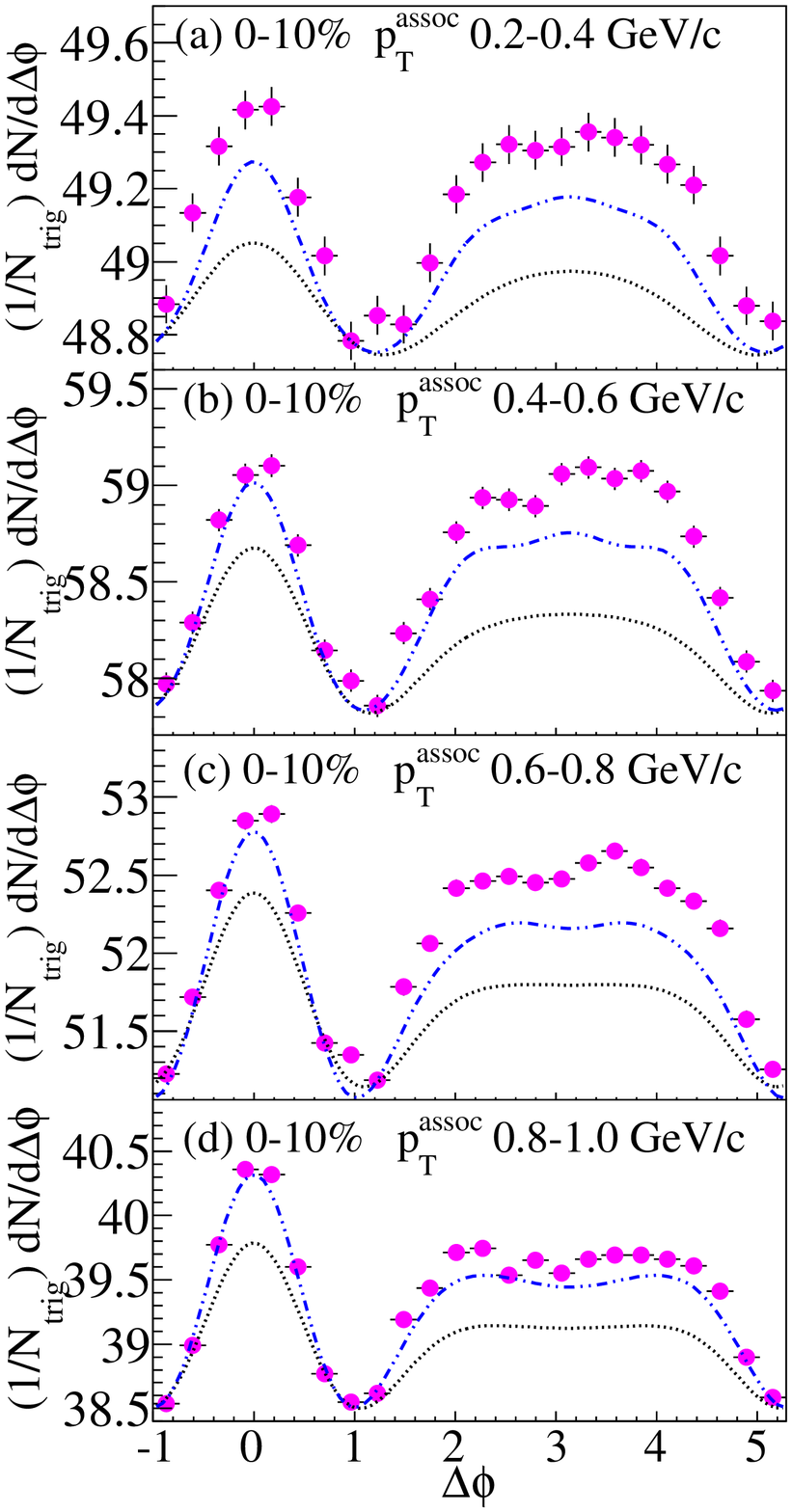}
\includegraphics[width=5.8cm]{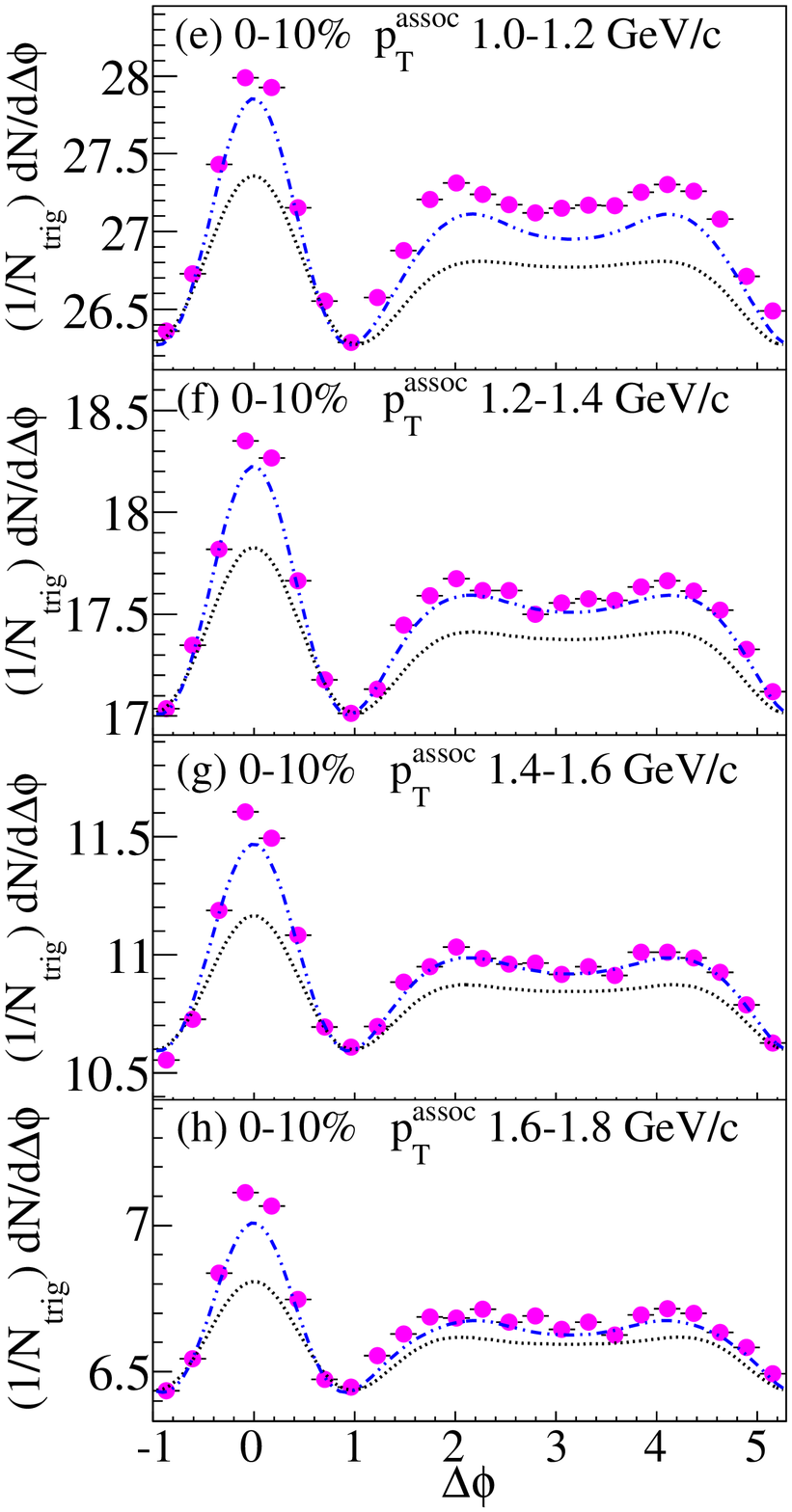}
\includegraphics[width=5.8cm]{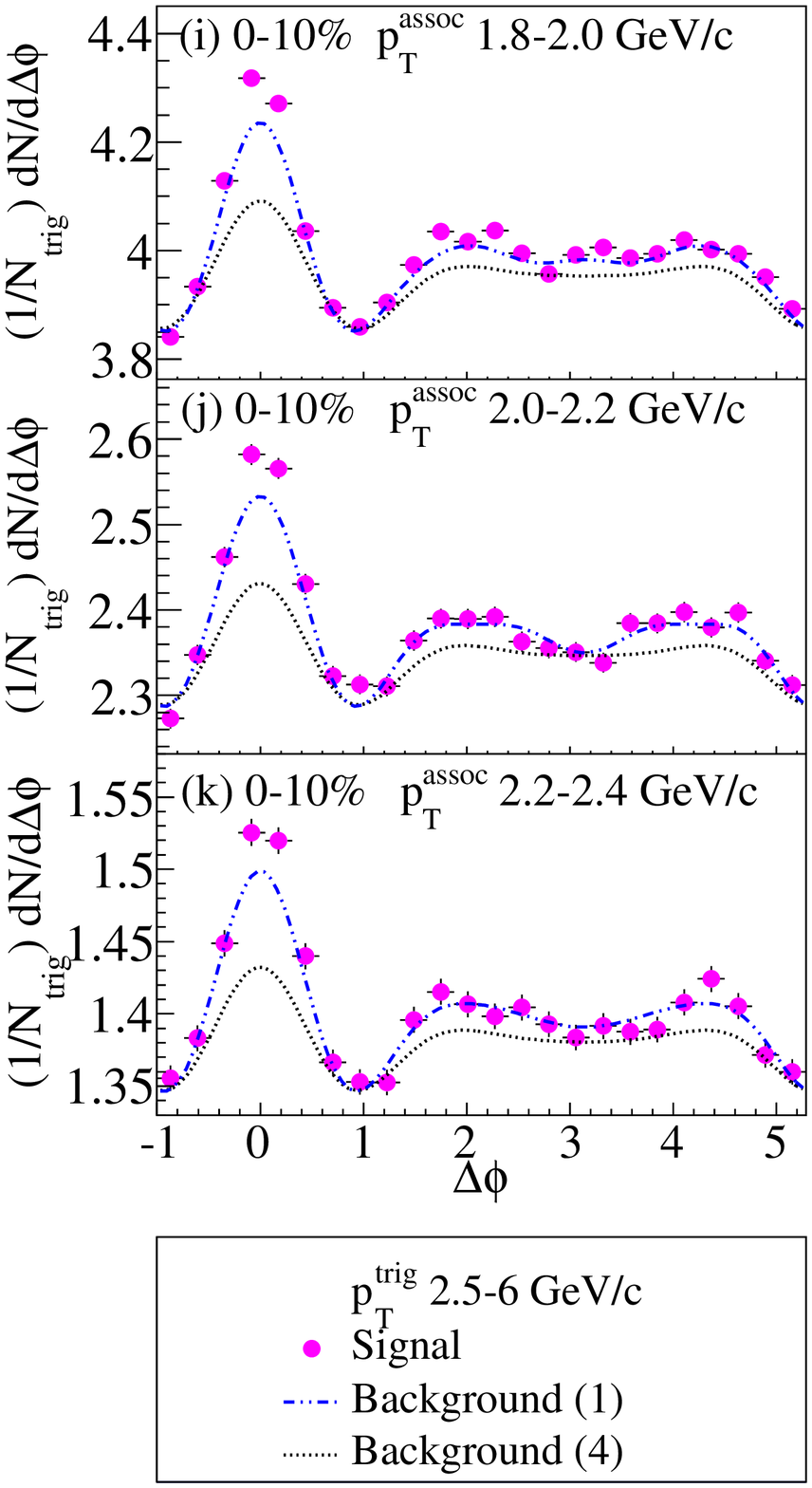}
\caption{(Color online) Dihadron correlation functions for different $p_{T}^{assoc}$ bins in 200 GeV/c
Au+Au collisions at 0-10\% centrality. From upper-left to bottom-right, the  $p_{T}^{assoc}$ are separated into
11 bins with 0.2 GeV/c bin width. Solid circles are raw signals, dash-dotted blue curves are the combinational
background (1) cases and dotted black curves are the combinational background (4) cases.}
\label{200GeV_dihadron}
\end{figure*}

\begin{figure*}[htbp]
\includegraphics[width=5.8cm]{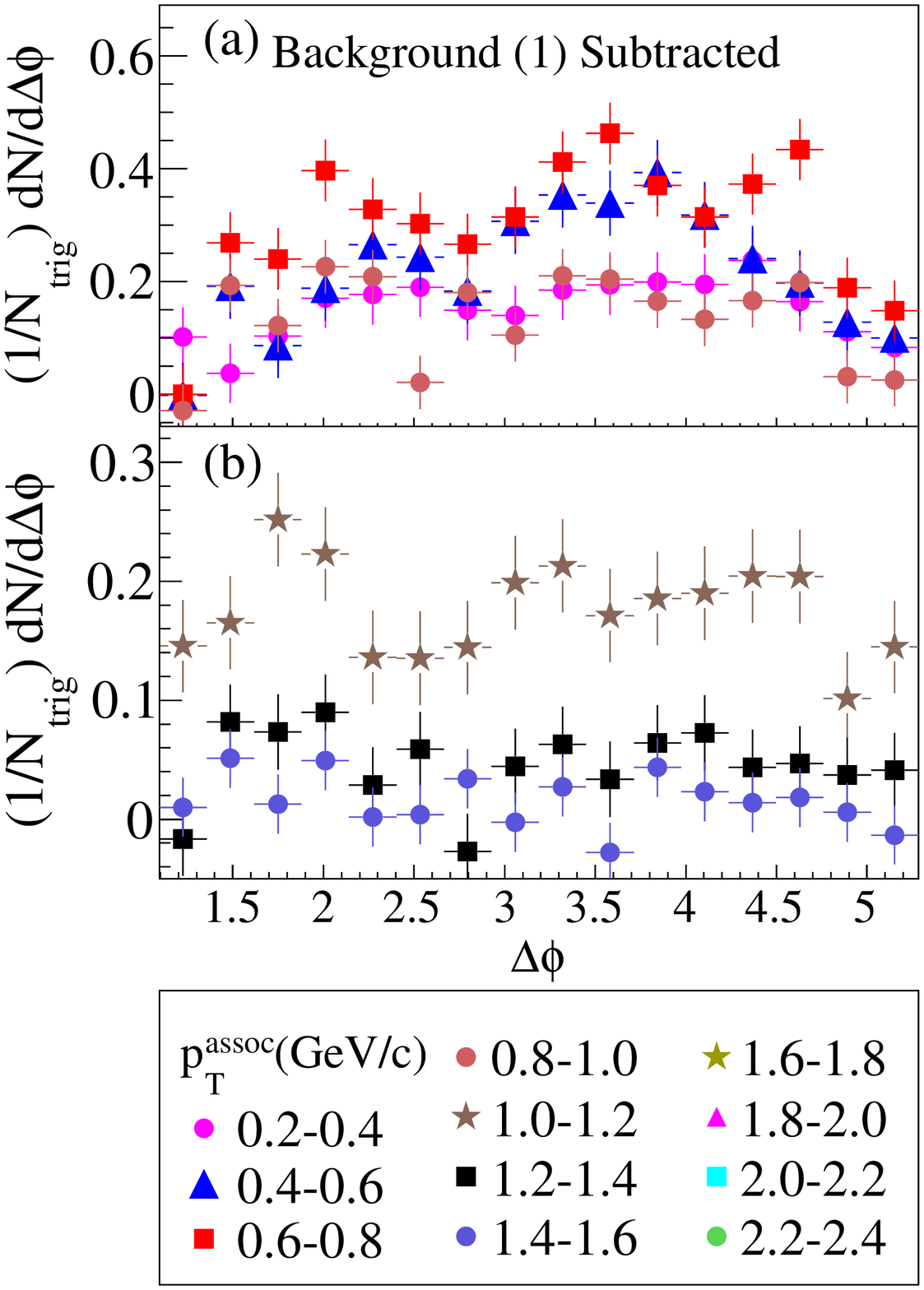}
\includegraphics[width=5.8cm]{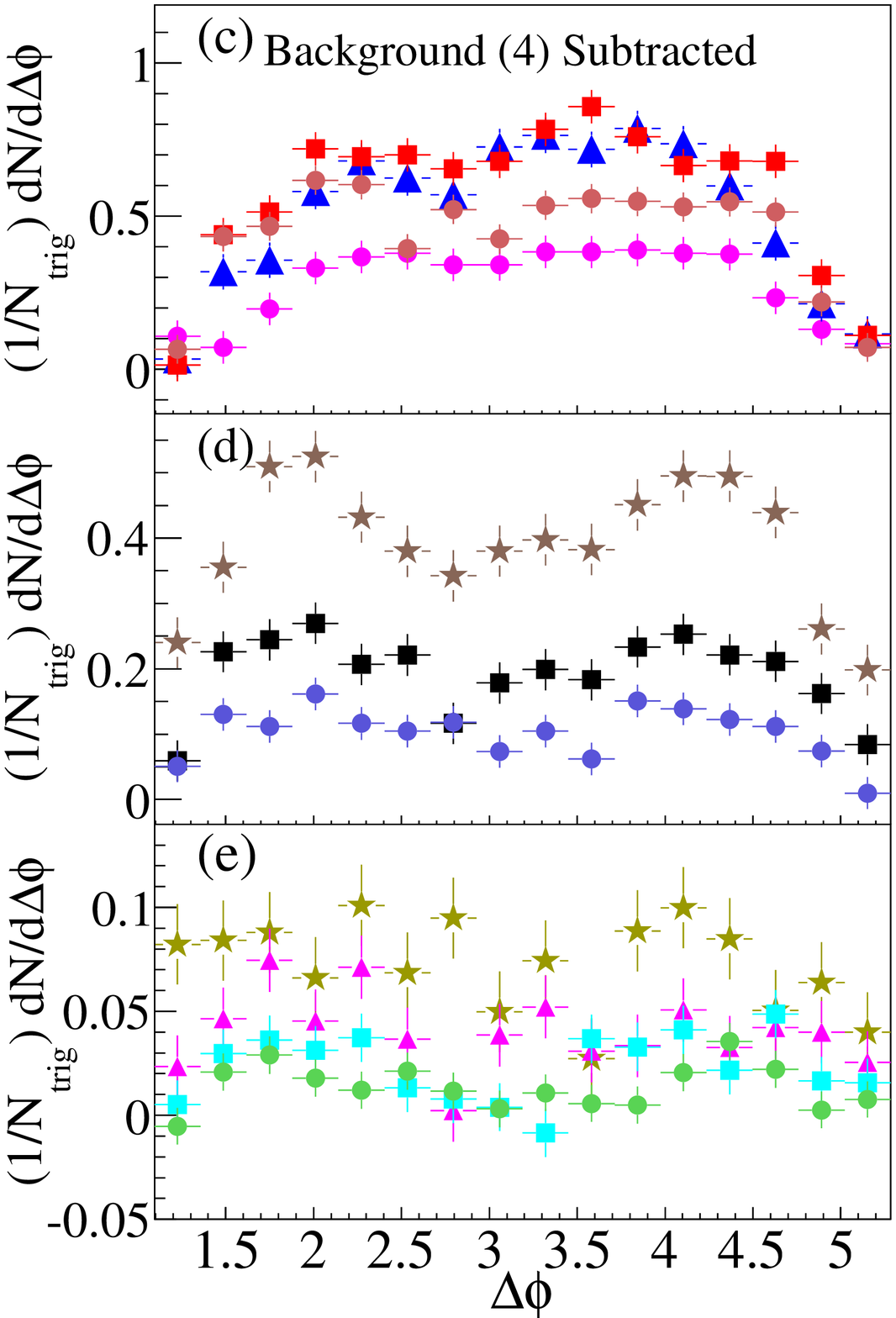}
\includegraphics[width=5.8cm]{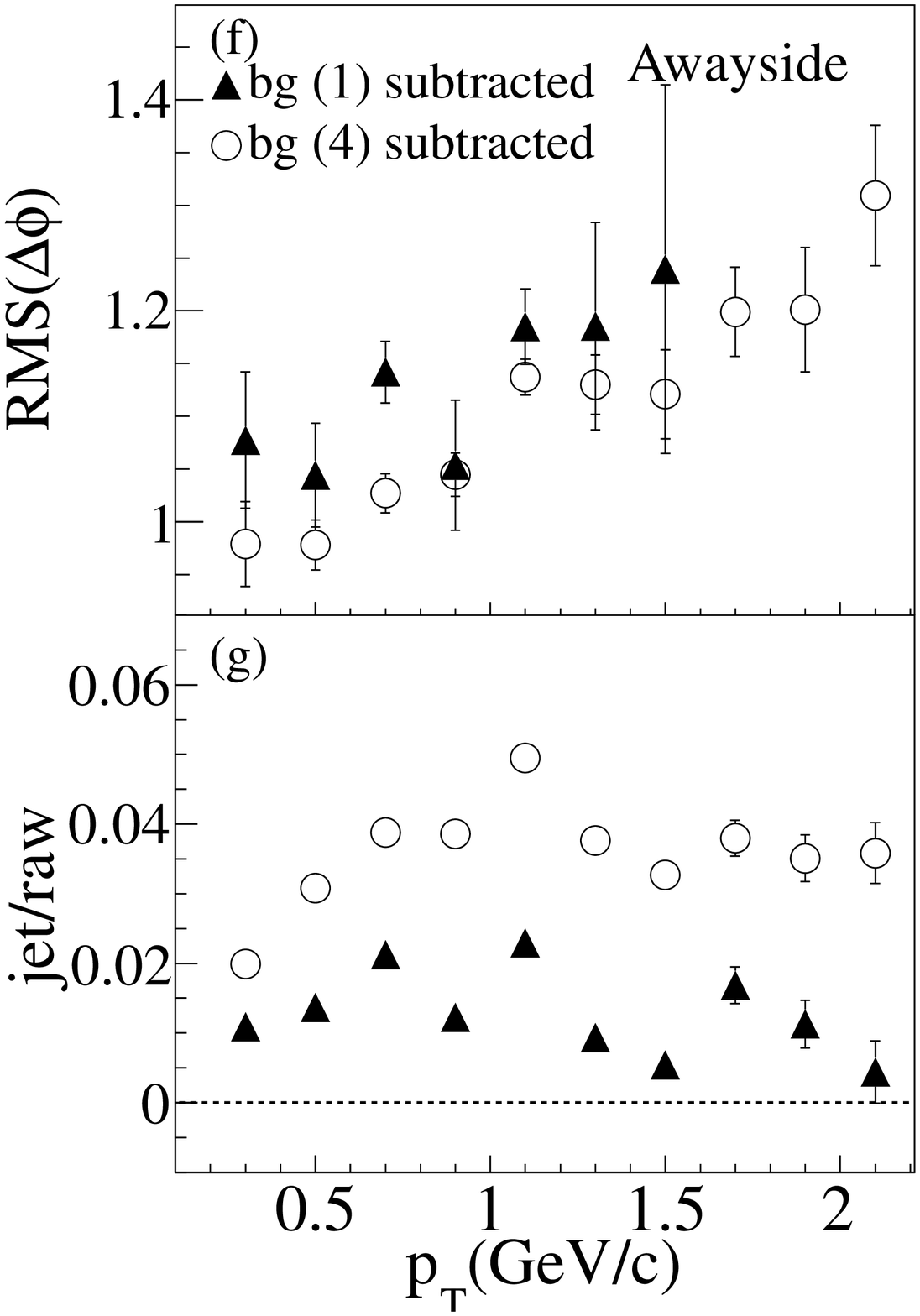}
\caption{(Color online) Panels (a), (b): background subtracted away-side signals for seven different
$p_{T}^{assoc}$ bins for background (1) case; Panels (c), (d), (e): away-side signals for twelve
different bins for background (4) case; Panels (f), (g): $p_{T}^{assoc}$ dependences
of RMS width and jet relative contribution.}
\label{200GeV_jet_real_9pT}
\end{figure*}

Since we have already observed the modification to the correlation
function by jet-medium interactions in Fig. 2 and 3, next we would like to focus central 0-10\% events for further
analysis since the modification is the most obvious there. We divide the whole $p_{T}^{assoc}$ range (from 0.2 GeV/c to 2.4 GeV/c) into much
smaller bins (with  a bin width of 0.2 GeV/c),  which are shown in Fig. 4.

In Fig. 4 , since background (1) almost overlaps with raw signal in high $p_{T}$ range
(larger than 1.6 GeV/c), we only pick out 7 $p_{T}$ bin results for background
(1) case. For background (2) case, all the real signals are extracted. They
are drawn in Fig. 5.

Panels (a) to (e) in Fig. 5 show evolution of background
subtracted dihadron correlations with the increase of
$p_{T}^{assoc}$. For background (1) case (panels (a), (b)), we can see
that the signal shape becomes flatter and flatter with
$p_{T}^{assoc}$. For background (4) case (panels (c) to (e)), a clear
evolution from a flat or seemingly single peak structure to a
double peak structure with $p_{T}^{assoc}$, because the
unsubstracted $\upsilon _{3}$ fluctuation and correlation remain
partly there. RMS and jet relative contribution are extracted from
the two panels, which are shown in the panels (f) and (g) in Fig. 5. The
$p_{T}^{assoc}$ dependence of RMS also shows the similar
evolutions of away-side signal shape for both cases. The
reasonable range of jet relative contribution is shown in the
panel (g). These results on the most central collisions give us
a more detailed picture that jets strongly interact with the
partonic medium.

\section{Summary}
In summary, we study dihadron azimuthal correlation functions  by a
multi-phase transport model. We obtain the harmonic flows with less nonflow effect and construct the combined harmonic
flow background using two formulas as reasonable upper and lower limits. Although the backgrounds calculated by two formulas differ in magnitude, the
physics information is quite similar for both cases.

Dihadron azimuthal correlations in 200 GeV/c Au+Au collisions with different centralities 0-10\%, 20-40\% and 50-80\%
are obtained in three $p_{T}^{assoc}$ bins. The evolution of real signal shape
and away-side jet relative contribution with the increase of $p_{T}^{assoc}$
and centrality is consistent with the fact that the high $p_{T}$
particles are strongly modified by the hot dense medium and the hot dense medium is less likely to
be generated or is weaker in more peripheral collisions.

More comprehensive study on $p_{T}^{assoc}$ dependence of jet-medium
contribution is done by dividing $p_{T}^{assoc}$ into more smaller bins with the width of
0.2 GeV/c. We can still observe the evolution from single gaussian shape
to flat or even double peak shape with the increase of $p_{T}^{assoc}$.

The jet contribution percentage in the raw dihadron correlation
function is very small. For the most central events (0-10\%), it is less than 5\%. Therefore, it is
a real challenge for extracting the jet-related signal experimentally,
since it requires a very good control of the harmonic flow background.

This work was supported in part by  the National Natural Science Foundation of
China under contract Nos. 11035009, 11220101005, 10979074,  11105207, 11175232
and the Knowledge Innovation Project of
the Chinese Academy of Sciences under Grant No. KJCX2-EW-N01.


\appendix
\section{List of values of flow parameters}
\begin{table*}
\caption{\label{tab:example}$\left \langle\upsilon_{1}^{trig}\right \rangle$  $\left \langle\upsilon_{1}^{assoc}\right \rangle$  $\left \langle\upsilon_{1}^{trig} \upsilon_{1}^{assoc} \right \rangle$}

\begin{ruledtabular}
\begin{tabular}{llllllllll}
   &  & 0-10\% &  &   & 20-40\% &  &  & 50-80\% &  \\
   $p_{T}^{assoc}$ range & $\left \langle\upsilon_{1}^{trig}\right \rangle$ & $\left \langle\upsilon_{1}^{assoc}\right \rangle$ & $\left \langle\upsilon_{1}^{trig} \upsilon_{1}^{assoc} \right \rangle$ 
   & $\left \langle\upsilon_{1}^{trig}\right \rangle$ & $\left \langle\upsilon_{1}^{assoc}\right \rangle$ & $\left \langle\upsilon_{1}^{trig} \upsilon_{1}^{assoc} \right \rangle$ 
   & $\left \langle\upsilon_{1}^{trig}\right \rangle$ & $\left \langle\upsilon_{1}^{assoc}\right \rangle$ & $\left \langle\upsilon_{1}^{trig} \upsilon_{1}^{assoc} \right \rangle$ 
\\

0.2-0.4 GeV/c &0.001837 &0.003891 &-0.000697 &0.001342 &0.001342 &-0.000942
&0.000993 &0.000993 &-0.001940\\
		0.4-0.6 GeV/c &0.003891 &0.004727 &-0.001506 &0.001394 &0.001394
		&-0.001624 &0.000642 &0.000642 &-0.002396\\
			0.6-0.8 GeV/c &0.004727 &0.003769 &-0.001587 &0.001312 &0.001312
			&-0.001456 &-0.000149 &-0.000149 &-0.002416\\
				0.8-1.0 GeV/c &0.003769 &0.001927 &-0.001085 &0.000702 &0.000702
				&-0.001577 &-0.000068 &-0.000068 &-0.002329\\
					1.0-1.2 GeV/c &0.001927 &-0.000342 &-0.000595 &0.000570 &0.000570
					&-0.001468 &0.000488 &0.000488 &-0.003082\\
						1.2-1.4 GeV/c &-0.000342 &-0.001999 &-0.000035 &-0.000051
						&-0.000051 &-0.000855 &-0.001384 &-0.001384 &-0.003353\\
							1.4-1.6 GeV/c &-0.001999 &-0.004586 &0.000807 &-0.000390
							&-0.000390 &-0.000036 &-0.003510 &-0.003510 &-0.003151\\
								1.6-1.8 GeV/c &-0.004586 &-0.007725 &0.001316 &-0.001960
								&-0.001960 &0.000864 &0.002034 &0.002034 &-0.003610\\
									1.8-2.0 GeV/c &-0.007725 &-0.010687 &0.002172 &-0.003233
									&-0.003233 &0.001257 &0.000679 &0.000679 &-0.003284\\
										2.0-2.2 GeV/c &-0.010687 &-0.012145 &0.003433 &-0.002566
										&-0.002566 &0.001859 &-0.002832 &-0.002832 &-0.003654\\
											2.2-2.4 GeV/c &-0.012145 &-0.016067 &0.003415 &-0.003811
											&-0.003811 &0.002671 &-0.005312 &-0.005312 &-0.004666\\
\end{tabular}

\end{ruledtabular}
\end{table*}

\begin{table*}
\caption{\label{tab:example}$\left \langle\upsilon_{2}^{trig}\right \rangle$  $\left \langle\upsilon_{2}^{assoc}\right \rangle$  $\left \langle\upsilon_{2}^{trig} \upsilon_{2}^{assoc} \right \rangle$}

\begin{ruledtabular}
\begin{tabular}{llllllllll}
   &  & 0-10\% &  &   & 20-40\% &  &  & 50-80\% &  \\
   $p_{T}^{assoc}$ range & $\left \langle\upsilon_{2}^{trig}\right \rangle$ & $\left \langle\upsilon_{2}^{assoc}\right \rangle$ & $\left \langle\upsilon_{2}^{trig} \upsilon_{2}^{assoc} \right \rangle$ 
   & $\left \langle\upsilon_{2}^{trig}\right \rangle$ & $\left \langle\upsilon_{2}^{assoc}\right \rangle$ & $\left \langle\upsilon_{2}^{trig} \upsilon_{2}^{assoc} \right \rangle$ 
   & $\left \langle\upsilon_{2}^{trig}\right \rangle$ & $\left \langle\upsilon_{2}^{assoc}\right \rangle$ & $\left \langle\upsilon_{2}^{trig} \upsilon_{2}^{assoc} \right \rangle$ 
\\
0.2-0.4 GeV/c &0.042004 &0.016994 &0.001957 &0.041964 &0.041964 &0.009125
&0.052629 &0.052629 &0.012003\\
		0.4-0.6 GeV/c &0.016994 &0.031763 &0.003030 &0.070599 &0.070599 &0.015442
		&0.081647 &0.081647 &0.018957\\
			0.6-0.8 GeV/c &0.031763 &0.044547 &0.004151 &0.095192 &0.095192
			&0.021016 &0.103296 &0.103296 &0.024413\\
				0.8-1.0 GeV/c &0.044547 &0.053954 &0.004852 &0.114463 &0.114463
				&0.025449 &0.120846 &0.120846 &0.028254\\
					1.0-1.2 GeV/c &0.053954 &0.060495 &0.005207 &0.129821 &0.129821
					&0.029124 &0.133234 &0.133234 &0.032168\\
						1.2-1.4 GeV/c &0.060495 &0.064775 &0.006021 &0.141029 &0.141029
						&0.031751 &0.143701 &0.143701 &0.035081\\
							1.4-1.6 GeV/c &0.064775 &0.067712 &0.006900 &0.149935 &0.149935
							&0.034043 &0.151426 &0.151426 &0.038436\\
								1.6-1.8 GeV/c &0.067712 &0.068329 &0.007220 &0.156176
								&0.156176 &0.035729 &0.152973 &0.152973 &0.038768\\
									1.8-2.0 GeV/c &0.068329 &0.068417 &0.007574 &0.160479
									&0.160479 &0.037796 &0.155515 &0.155515 &0.040527\\
										2.0-2.2 GeV/c &0.068417 &0.066030 &0.007593 &0.163173
										&0.163173 &0.038348 &0.159956 &0.159956 &0.043669\\
											2.2-2.4 GeV/c &0.066030 &0.064178 &0.008366 &0.163583
											&0.163583 &0.038770 &0.160305 &0.160305 &0.043453\\
\end{tabular}

\end{ruledtabular}
\end{table*}

\begin{table*}
\caption{\label{tab:example}$\left \langle\upsilon_{3}^{trig}\right \rangle$  $\left \langle\upsilon_{3}^{assoc}\right \rangle$  $\left \langle\upsilon_{3}^{trig} \upsilon_{3}^{assoc} \right \rangle$}
\begin{ruledtabular}
\begin{tabular}{llllllllll}
   &  & 0-10\% &  &   & 20-40\% &  &  & 50-80\% &  \\
   $p_{T}^{assoc}$ range & $\left \langle\upsilon_{3}^{trig}\right \rangle$ & $\left \langle\upsilon_{3}^{assoc}\right \rangle$ & $\left \langle\upsilon_{3}^{trig} \upsilon_{3}^{assoc} \right \rangle$ 
   & $\left \langle\upsilon_{3}^{trig}\right \rangle$ & $\left \langle\upsilon_{3}^{assoc}\right \rangle$ & $\left \langle\upsilon_{3}^{trig} \upsilon_{3}^{assoc} \right \rangle$ 
   & $\left \langle\upsilon_{3}^{trig}\right \rangle$ & $\left \langle\upsilon_{3}^{assoc}\right \rangle$ & $\left \langle\upsilon_{3}^{trig} \upsilon_{3}^{assoc} \right \rangle$ 
\\
0.2-0.4 GeV/c &0.030201 &0.004909 &0.001118 &0.010104 &0.010104 &0.001703
&0.012910 &0.012910 &0.002508\\
		0.4-0.6 GeV/c &0.004909 &0.015469 &0.002593 &0.021127 &0.021127 &0.003401
		&0.019854 &0.019854 &0.003226\\
			0.6-0.8 GeV/c &0.015469 &0.028102 &0.004178 &0.032916 &0.032916
			&0.005453 &0.027854 &0.027854 &0.004895\\
				0.8-1.0 GeV/c &0.028102 &0.040093 &0.006086 &0.043845 &0.043845
				&0.007427 &0.033936 &0.033936 &0.006076\\
					1.0-1.2 GeV/c &0.040093 &0.050931 &0.007859 &0.053737 &0.053737
					&0.008962 &0.039496 &0.039496 &0.006865\\
						1.2-1.4 GeV/c &0.050931 &0.059100 &0.008760 &0.061137 &0.061137
						&0.010424 &0.045423 &0.045423 &0.008008\\
							1.4-1.6 GeV/c &0.059100 &0.066054 &0.009964 &0.068773 &0.068773
							&0.011919 &0.048002 &0.048002 &0.008809\\
								1.6-1.8 GeV/c &0.066054 &0.071490 &0.010496 &0.074461
								&0.074461 &0.013307 &0.051733 &0.051733 &0.010559\\
									1.8-2.0 GeV/c &0.071490 &0.075416 &0.011367 &0.077711
									&0.077711 &0.013558 &0.054062 &0.054062 &0.010938\\
										2.0-2.2 GeV/c &0.075416 &0.076608 &0.012745 &0.082178
										&0.082178 &0.014395 &0.058495 &0.058495 &0.008238\\
											2.2-2.4 GeV/c &0.076608 &0.078378 &0.013210 &0.085642
											&0.085642 &0.014619 &0.061674 &0.061674 &0.014042\\
\end{tabular}

\end{ruledtabular}
\end{table*}

\begin{table*}
\caption{\label{tab:example}$\left \langle\upsilon_{4}^{trig}\right \rangle$  $\left \langle\upsilon_{4}^{assoc}\right \rangle$  $\left \langle\upsilon_{4}^{trig} \upsilon_{4}^{assoc} \right \rangle$}
\begin{ruledtabular}
\begin{tabular}{llllllllll}
   &  & 0-10\% &  &   & 20-40\% &  &  & 50-80\% &  \\
   $p_{T}^{assoc}$ range & $\left \langle\upsilon_{4}^{trig}\right \rangle$ & $\left \langle\upsilon_{4}^{assoc}\right \rangle$ & $\left \langle\upsilon_{4}^{trig} \upsilon_{4}^{assoc} \right \rangle$ 
   & $\left \langle\upsilon_{4}^{trig}\right \rangle$ & $\left \langle\upsilon_{4}^{assoc}\right \rangle$ & $\left \langle\upsilon_{4}^{trig} \upsilon_{4}^{assoc} \right \rangle$ 
   & $\left \langle\upsilon_{4}^{trig}\right \rangle$ & $\left \langle\upsilon_{4}^{assoc}\right \rangle$ & $\left \langle\upsilon_{4}^{trig} \upsilon_{4}^{assoc} \right \rangle$ 
\\
0.2-0.4 GeV/c &0.015998 &0.000948 &0.000221 &0.001625 &0.001625 &0.000085
&-0.001360 &-0.001360 &0.000677\\
		0.4-0.6 GeV/c &0.000948 &0.005678 &0.000723 &0.004099 &0.004099 &0.000683
		&-0.002469 &-0.002469 &0.000986\\
			0.6-0.8 GeV/c &0.005678 &0.012493 &0.001632 &0.007566 &0.007566
			&0.001628 &-0.002682 &-0.002682 &0.001783\\
				0.8-1.0 GeV/c &0.012493 &0.020305 &0.002501 &0.011671 &0.011671
				&0.002133 &-0.003266 &-0.003266 &0.002189\\
					1.0-1.2 GeV/c &0.020305 &0.028170 &0.003525 &0.015414 &0.015414
					&0.003059 &-0.002346 &-0.002346 &0.003384\\
						1.2-1.4 GeV/c &0.028170 &0.035320 &0.004594 &0.018721 &0.018721
						&0.004194 &-0.002007 &-0.002007 &0.002937\\
							1.4-1.6 GeV/c &0.035320 &0.042858 &0.005170 &0.021235 &0.021235
							&0.004439 &-0.002812 &-0.002812 &0.002656\\
								1.6-1.8 GeV/c &0.042858 &0.048819 &0.005748 &0.025388
								&0.025388 &0.005695 &-0.004228 &-0.004228 &0.001316\\
									1.8-2.0 GeV/c &0.048819 &0.053568 &0.006991 &0.028373
									&0.028373 &0.006568 &-0.005650 &-0.005650 &0.002296\\
										2.0-2.2 GeV/c &0.053568 &0.056891 &0.006922 &0.029028
										&0.029028 &0.006538 &-0.002869 &-0.002869 &0.005727\\
											2.2-2.4 GeV/c &0.056891 &0.058573 &0.007499 &0.031020
											&0.031020 &0.005983 &-0.000363 &-0.000363 &0.005555\\
\end{tabular}

\end{ruledtabular}
\end{table*}

\begin{table*}
\caption{\label{tab:example}$\left \langle\upsilon_{5}^{trig}\right \rangle$  $\left \langle\upsilon_{5}^{assoc}\right \rangle$  $\left \langle\upsilon_{5}^{trig} \upsilon_{5}^{assoc} \right \rangle$}
\begin{ruledtabular}
\begin{tabular}{llllllllll}
   &  & 0-10\% &  &   & 20-40\% &  &  & 50-80\% &  \\
   $p_{T}^{assoc}$ range & $\left \langle\upsilon_{5}^{trig}\right \rangle$ & $\left \langle\upsilon_{5}^{assoc}\right \rangle$ & $\left \langle\upsilon_{5}^{trig} \upsilon_{5}^{assoc} \right \rangle$ 
   & $\left \langle\upsilon_{5}^{trig}\right \rangle$ & $\left \langle\upsilon_{5}^{assoc}\right \rangle$ & $\left \langle\upsilon_{5}^{trig} \upsilon_{5}^{assoc} \right \rangle$ 
   & $\left \langle\upsilon_{5}^{trig}\right \rangle$ & $\left \langle\upsilon_{5}^{assoc}\right \rangle$ & $\left \langle\upsilon_{5}^{trig} \upsilon_{5}^{assoc} \right \rangle$ 
\\
0.2-0.4 GeV/c &0.005908 &0.000163 &0.000072 &-0.000210 &-0.000210 &0.000189
&-0.001752 &-0.001752 &-0.000043\\
		0.4-0.6 GeV/c &0.000163 &0.001217 &0.000029 &-0.000709 &-0.000709
		&0.000279 &-0.003521 &-0.003521 &-0.000147\\
			0.6-0.8 GeV/c &0.001217 &0.003841 &0.000401 &-0.000459 &-0.000459
			&0.000549 &-0.004073 &-0.004073 &0.000861\\
				0.8-1.0 GeV/c &0.003841 &0.006993 &0.000565 &-0.000890 &-0.000890
				&0.000854 &-0.004225 &-0.004225 &0.000594\\
					1.0-1.2 GeV/c &0.006993 &0.010640 &0.001154 &-0.000437 &-0.000437
					&0.000669 &-0.005513 &-0.005513 &-0.000439\\
						1.2-1.4 GeV/c &0.010640 &0.014408 &0.001556 &-0.000367 &-0.000367
						&0.000239 &-0.005430 &-0.005430 &0.001545\\
							1.4-1.6 GeV/c &0.014408 &0.018030 &0.001783 &0.000923 &0.000923
							&0.000809 &-0.006749 &-0.006749 &0.001939\\
								1.6-1.8 GeV/c &0.018030 &0.020953 &0.002676 &0.002621
								&0.002621 &0.001746 &-0.007577 &-0.007577 &0.002326\\
									1.8-2.0 GeV/c &0.020953 &0.023771 &0.002333 &0.001566
									&0.001566 &0.001071 &-0.005370 &-0.005370 &0.001255\\
										2.0-2.2 GeV/c &0.023771 &0.027007 &0.003120 &0.001490
										&0.001490 &0.002077 &-0.011794 &-0.011794 &0.001494\\
											2.2-2.4 GeV/c &0.027007 &0.030243 &0.002670 &0.003509
											&0.003509 &0.002326 &-0.011982 &-0.011982 &0.001287\\
\end{tabular}

\end{ruledtabular}
\end{table*}

\begin{table*}
\caption{\label{tab:example}$\left \langle\upsilon_{6}^{trig}\right \rangle$  $\left \langle\upsilon_{6}^{assoc}\right \rangle$  $\left \langle\upsilon_{6}^{trig} \upsilon_{6}^{assoc} \right \rangle$}
\begin{ruledtabular}
\begin{tabular}{llllllllll}
   &  & 0-10\% &  &   & 20-40\% &  &  & 50-80\% &  \\
   $p_{T}^{assoc}$ range & $\left \langle\upsilon_{6}^{trig}\right \rangle$ & $\left \langle\upsilon_{6}^{assoc}\right \rangle$ & $\left \langle\upsilon_{6}^{trig} \upsilon_{6}^{assoc} \right \rangle$ 
   & $\left \langle\upsilon_{6}^{trig}\right \rangle$ & $\left \langle\upsilon_{6}^{assoc}\right \rangle$ & $\left \langle\upsilon_{6}^{trig} \upsilon_{6}^{assoc} \right \rangle$ 
   & $\left \langle\upsilon_{6}^{trig}\right \rangle$ & $\left \langle\upsilon_{6}^{assoc}\right \rangle$ & $\left \langle\upsilon_{6}^{trig} \upsilon_{6}^{assoc} \right \rangle$ 
\\
0.2-0.4 GeV/c &0.001174 &-0.000231 &0.000120 &0.000010 &0.000010 &-0.000044
&0.000120 &0.000120 &-0.000396\\
		0.4-0.6 GeV/c &-0.000231 &0.000030 &0.000215 &-0.000491 &-0.000491
		&-0.000084 &-0.001390 &-0.001390 &-0.000110\\
			0.6-0.8 GeV/c &0.000030 &0.000156 &-0.000083 &-0.001263 &-0.001263
			&-0.000056 &-0.001474 &-0.001474 &-0.000271\\
				0.8-1.0 GeV/c &0.000156 &0.001058 &0.000036 &-0.002195 &-0.002195
				&0.000216 &0.000550 &0.000550 &0.000498\\
					1.0-1.2 GeV/c &0.001058 &0.002324 &0.000420 &-0.002858 &-0.002858
					&0.000355 &-0.000728 &-0.000728 &0.000465\\
						1.2-1.4 GeV/c &0.002324 &0.003721 &0.000349 &-0.003420 &-0.003420
						&0.000420 &-0.002572 &-0.002572 &-0.000688\\
							1.4-1.6 GeV/c &0.003721 &0.004106 &0.000447 &-0.003858
							&-0.003858 &0.000966 &-0.002692 &-0.002692 &0.000093\\
								1.6-1.8 GeV/c &0.004106 &0.006095 &0.000886 &-0.005499
								&-0.005499 &0.000622 &-0.002420 &-0.002420 &0.002307\\
									1.8-2.0 GeV/c &0.006095 &0.007962 &0.001136 &-0.004095
									&-0.004095 &0.000763 &-0.002748 &-0.002748 &0.001520\\
										2.0-2.2 GeV/c &0.007962 &0.007611 &-0.000213 &-0.001545
										&-0.001545 &0.000804 &0.002193 &0.002193 &-0.000750\\
											2.2-2.4 GeV/c &0.007611 &0.008142 &0.000370 &-0.001668
											&-0.001668 &0.000705 &0.003055 &0.003055 &0.000971\\
\end{tabular}

\end{ruledtabular}
\end{table*}

\end{document}